\documentclass[prl,floatfix,showpacs,amsmath,twocolumn]{revtex4}
\usepackage{amssymb}
\usepackage{graphicx}
\usepackage{graphics}

\newcommand{\bea}{\begin{eqnarray}}
\newcommand{\eea}{\end{eqnarray}}
\def\bi{\begin{itemize}}
\def\ei{\end{itemize}}
\def\bc{\begin{center}}
\def\ec{\end{center}}

\newcommand*{\cS}{\mathcal{S}}
\newcommand*{\QBER}{\mathrm{QBER}}
\newcommand*{\cD}{\mathcal{D}}

\def\<{\langle}
\def\>{\rangle}

\def\opone{\leavevmode\hbox{\small1\kern-3.8pt\normalsize1}}

\def\ket#1{|#1\rangle}

\newcommand{\one}{\mbox{$1 \hspace{-1.0mm}  {\bf l}$}}
\def\tr{\mathrm{tr}}
\def\ket#1{\left| #1\right>}
\def\bra#1{\left< #1\right|}

\newcommand{\proj}[1]{\ket{#1}\bra{#1}}

\newcommand*{\eps}{\varepsilon}

\newcommand{\cancel}[1]{}

\begin{document}
\title{Lower and upper bounds on the secret key rate for quantum key distribution protocols using one--way classical communication}

\author{B. Kraus$^{1}$, N. Gisin$^{1}$ and R. Renner$^{2}$}
\affiliation{$^1$ Group of Applied Physics, University of Geneva,
CH--$1211$ Geneve $4$, Switzerland, \\
$^2$ Computer Science Department, ETH--Z\"urich, Switzerland}

\begin{abstract}
  We investigate a general class of quantum key distribution (QKD) protocols using one-way
  classical communication. We show that full security can be proven by considering only collective attacks.
  We derive computable lower and upper bounds on the secret key rate of those QKD protocol
  involving only entropies of two--qubit density operators. As an illustration of our results,
  we determine new bounds for the BB84, the six-state, and the B92 protocol. We show that in all these cases
  the first classical processing that the legitimate partners
  should apply consists in adding noise. This is precisely why any entanglement based proof would generally fail
  here.
\end{abstract}

\pacs{03.67.Dd,03.67.-a}

\maketitle

Quantum cryptography, the art of exploiting quantum physics to
defeat any possible eavesdropper, has rapidly grown over the last
decade from the level of a nice idea into an entire branch of
physics \cite{GiRi02}. Indeed, first commercial equipment are
already offered \cite{Idq}.

A generic QKD protocol can be divided into two parts: I)
Distribution of quantum information and measurement II) Classical
part consisting out of parameter estimation and classical
post-processing (CPP). To implement the quantum part of the
protocol, the two legitimate persons, Alice ($A$) and Bob ($B$),
agree on some encoding/decoding procedure \cite{footnote2}. We
denote by ${\cal
  S}_0=\{\ket{\phi^0_j}\}_{j\in J}$ and ${\cal
  S}_1=\{\ket{\phi^1_j}\}_{j\in J}$, where $J=\{1,\ldots, m\}$, the
sets of states used to encode the bit value $0$, $1$, resp..
First, $A$ sends $n$ qubits prepared at random in the state
$\ket{\phi^{i_1}_{j_1}}\otimes\ldots \otimes
\ket{\phi^{i_n}_{j_n}}\equiv \ket{\phi^{\bold{i}}_{\bold{j}}}$ to
$B$ \cite{footnote5}. The adversary, Eve ($E$), interacts now with
all the qubits sent by $A$. She applies a unitary transformation
to all those qubits and an ancilla in the state $\ket{0}$
\cite{footnote6}. The state $E$ and $B$ share then is given by
$\ket{\Phi^{\bold{i}}_{\bold{j}}}_{BE}\equiv {\cal
  U}_{BE}\ket{\phi^{\bold{i}}_{\bold{j}}}_B \ket{0}_E$. Next, $B$
applies some filtering operation and measures his qubits in the
$z$--basis \cite{footnote61}. $A$ and $B$ compare publicly which
encoding/decoding operation they used and keep only those pairs of
qubits where they were compatible (sifting). The state describing
$E$'s system is $
\ket{\Phi^{\bold{i},\bold{k}}_{\bold{j}}}_{E}\equiv \bra{\bold{k}}
B_{\bold j}{\cal U}_{BE}\ket{\phi^{\bold{i}}_{\bold{j}}}_B
\ket{0}_E,$ where we denoted by $B_{\bold j}$ the filtering
operation used by $B$ and by ${\bold k}$ his $z$--measurement
outcome \cite{footnote0}. $A$ and $B$ compare now publicly some of
their measurement outcomes to estimate the quantum bit error rate
(QBER).

The security of the protocol relies on the fact that $E$, trying
to gain information about the bit values, introduces some error
due to the laws of quantum mechanics. However, any realistic
channel used by $A$ and $B$ is noisy, i.e. $QBER>0$. In order to
ensure that the protocol is secure one must assume that all the
noise (estimated by $A$ and $B$) is due to an unlimited
eavesdropping attack, a coherent attack \cite{footnote7,
footnote31}. $A$ and $B$ know how to counter such an adversary:
they apply a CPP, consisting out of error correction (EC) and
privacy amplification (PA). This general principle leaves a
central question open: How much error can be tolerated in order to
be able to distill a secret key? This is precisely what we
concentrate on in this paper.

Previous security proofs are based on the following observations
\cite{ShPr00,Lo,TaKo03}. Instead of preparing a system and then
sending it to $B$, $A$ can equivalently prepare $B$'s system at a
distance by using an entangled state (entanglement--based scheme).
If $A$ and $B$ could purify their state to singlets, their systems
cannot be entangled to $E$. The essential feature can be carried
out processing only classical data, leading to perfectly
correlated data.

We present here a different, not on entanglement based, kind of
security proof for a class of QKD protocols including the BB84,
the 6--state, and the B92 protocol \cite{BB84,BeGi99,Be92}. First
of all, we determine the state shared by $A$ and $B$ (using the
entanglement--based scheme) after a general eavesdropping attack.
Then we analyze the classical part of the protocol, i.e.,
parameter estimation and CPP, for the case of one-way
communication. We present a new formula for the secret key length.
Then we derive a lower bound on the secret key rate involving only
entropies of two--qubit density operators. We also present an
upper bound on the secret key rate. At the end we illustrate our
results by determining new values for the lower bounds for the
BB84, the $6$--state, and the B92 protocol. These new bounds are
generally stronger than those achievable with entanglement--based
security proofs.


To study the entanglement based scheme we use the same notation as
before and define the encoding operators
$A_j=\ket{0}\bra{(\phi^0_j)^\ast}+\ket{1}\bra{(\phi^1_j)^\ast}$
and the decoding operators
$B_j=\ket{0}\bra{\hat{\phi}_j^1}+\ket{1}\bra{\hat{\phi}_j^0}$,
where $\ket{\hat{\phi}_j^i}$ denotes the orthogonal state to
$\ket{\phi_j^i}$ and $\ket{(\phi_j^i)^\ast}$ denotes the complex
conjugate of $\ket{\phi_j^i}$ in the computational basis for
$i=0,1$ and $j\in J$. Note that those operators are not
necessarily unitary, e.g. for the B92 protocol. After applying one
of those filtering operations $A$ and $B$ measure in the
$z$--basis, associating to the outcome the bit values $0$ or $1$.
Using the fact that $A^T\otimes \one \ket{\Phi^+}=\one\otimes A
\ket{\Phi^+}$ for any operator $A$ and $\ket{\Phi^+}=1/\sqrt{2}
(\ket{00}+\ket{11})$ and that the operators applied on $A$'s
systems commute with the operator applied by $E$ it is easy to
verify that
$\ket{\Phi^{\bold{i},\bold{k}}_{\bold{j}}}_{E}=_A\bra{\bold{i}}_B\bra{\bold{k}}
\Phi_{\bold{j}}\rangle_{ABE},$ where $\ket{\Phi_{\bold{j}}}_{ABE}=
A_{\bold{j}}\otimes B_{\bold{j}} {\cal
U}_{EB}\ket{\Phi^+}^{\otimes n}_{AB}\ket{0}_E$.

To account for all the different realizations (${\bold{j}}$) we
introduce a new system $R_1$ and define the state $
\ket{\chi_0}_{ABER}=\sum_{\bold{j}} \frac{1}{\sqrt{p_{\bold{j}}}}
\ket{\Phi_{\bold{j}}}_{ABE}\ket{\bold{j}}_{R_1},$ with $p_j$
determining the probability with which $A$ and $B$ decide to keep
the systems in case they used the operators $A_{\bold{j}},
B_{\bold{j}}$. Now, first of all $R_1$ measures and obtains the
outcome $\bold{j}$. The state shared by $A$, $B$, and $E$ is then
$\ket{\Phi_{\bold{j}}}_{ABE}$.

Let us now introduce an equivalent protocol where $A$ and $B$
additionally apply the following operations \cite{footnote10}: (I)
$A$ and $B$ apply both the same unitary transformation, $U_{{\bold
l}^\prime}$ chosen for each qubit at random among
$U_1=\one,U_2=\sigma_z$, with $\sigma_z$ one of the Pauli
operators. The equivalence to the previous protocol is due to the
fact that the state describing $E$'s system is not changed. (II)
$A$ and $B$ can decide to flip their bit values (both at the same
time). We combine the first two possible operations. The operator
$O_{l_i}$ denotes a unitary operator of the form $U_{l^\prime_i}
V_{l^{\prime \prime}_i}$, for
$l^\prime_i,l^{\prime\prime}_i\in\{1,2\}$, and $V_1=\one,
V_2=\sigma_x$. Since we assume that both apply the same operation,
they need to communicate classically. This exchanged classical
information will be denoted by ${\bold l}$. (III) $A$ and $B$ are
also free to permute their qubits/bit. Obviously, they have to use
the same permutation operators, $P_{\bold m}$. The classical
information which has to be exchanged is denoted by ${\bold m}$.

We introduce now two random number generators, $R_2$ and $R_3$,
which account respectively for the operators, $O_{\bold{l}}$ and
$P_{\bold m}$. The state describing all the systems is $
\ket{\chi}_{ABER_1R_2R_3}=\sum_{\bold{j},\bold{l},\bold{m}}
\frac{1}{\sqrt{p_{\bold{j}}}}
\ket{\Phi_{\bold{j},\bold{l},\bold{m}}}_{ABE}\ket{\bold{j}}_{R_1}\ket{\bold{l}}_{R_2}\ket{\bold{m}}_{R_3}$,
with $\ket{\Phi_{\bold{j},\bold{l},\bold{m}}}_{ABE}= P_m
O_{\bold{l}} A_{\bold{j}}\otimes P_m O_{\bold{l}} B_{\bold{j}}
{\cal U}_{EB}\ket{\Phi^+}^{\otimes n}_{AB}\ket{0}_E,$ the state
shared by $A$, $B$, and $E$ for the particular realization
$(\bold{j},\bold{l},\bold{m})$.

Let us now relax the assumptions about $E$. We provide $E$ with
all the systems $R_1,R_2,R_3$. Since she can measure the $R$
systems ending up in the same situation as before, we clearly
provide her with at least as much power as she had before. The
state $A$ and $B$ share is given by the partial trace of the state
$\ket{\chi}_{ABER_1R_2R_3}$ over $E,R_1,R_2,R_3$. We find $
\rho_{AB}^n={\cal P}_S \{{\cal D}^{\otimes n}_2 [{\cal D}^{\otimes
n}_1(\rho^0_{AB})]\}.$ Here the normalized state
$\rho^0_{AB}=\tr_E(P_{\ket{\psi_0}})$ with $\ket{\psi_0}={\cal
U}_{EB} \ket{\Phi^+}^{\otimes n}_{AB}\ket{0}_E$ and ${\cal P}_S$
the completely positive map (CPM) symmetrizing the state with
respect to all qubit pairs \cite{footnote9}. The CPM ${\cal D}_1$
is entirely defined by the protocol and is given by ${\cal
D}_1(\rho)=\sum_{\bold{j}} \frac{1}{p_{\bold{j}}}
A_{\bold{j}}\otimes B_{\bold{j}} (\rho)
A_{\bold{j}}^\dagger\otimes B_{\bold{j}}^\dagger$. ${\cal D}_2$ is
independent of the protocol, and is defined as ${\cal
D}_2(\rho)=\sum_{\bold{l}} O_\bold{l} \otimes O_\bold{l} (\rho)
O_\bold{l}^\dagger \otimes O_\bold{l}^\dagger$, i.e. the
depolarization map transforming any two--qubit state into a Bell
diagonal state. This implies that the density operator $A$ and $B$
share, before their measurement in the $z$--basis, has, for any
protocol the simple form \bea \label{rhogen} \rho_{AB}^n=\sum
\lambda_{n_1,n_2,n_3,n_4} {\cal P}_S (P_{\ket{\Phi_1}}^{\otimes
n_1}\otimes P_{\ket{\Phi_2}}^{\otimes n_2}\otimes
P_{\ket{\Phi_3}}^{\otimes n_3}\otimes P_{\ket{\Phi_4}}^{\otimes
n_4}). \eea Here, the sum is performed such that
$n_4=n-n_1-n_2-n_3$, with $n_i\geq 0$. The states
$\ket{\Phi_{1/2}}=1/\sqrt{2}(\ket{00}\pm \ket{11})$ and
$\ket{\Phi_{3/4}} =1/\sqrt{2}(\ket{10}\pm \ket{01})$ denote the
Bell basis. Note that this state is separable with respect to the
different qubit pairs. Note further that this result (Eq.
(\ref{rhogen})) is independent of the CPP, thus, it can also be
used in order to investigate any protocol employing two--way CPP.

The CPM ${\cal D}_2$ does not depend on the protocol and is only
due to the operations $O_{\bold l}$. In principle, $A$ and $B$ can
apply (independently) any unitary transformations of the sort
$e^{i \theta \sigma_z}$ to their qubits before they measure them
in the $z$--basis.  The state describing $E$'s system would then
be, up to a global phase, equivalent to
$\ket{\Phi^{\bold{i},\bold{k}}_{\bold{j}}}_{E}$. This can be also
seen as follows:  If the basis (say the $z$--basis) in which a
certain state, $\rho$ is measured is known then we can define a
set of operators which are {\it in the measurement basis
reducible} to $\rho$. Any state of the form $\rho^\prime=\sum_i
p_i O_i\otimes \one \rho O_i^\dagger \otimes \one$, with $p_i\geq
0, \sum_i p_i=1$ and unitary operators $O_i$ diagonal in the
measurement basis, i.e. $O_A\ket{i}=\lambda_i\ket{i}$, with $\mid
\lambda_i\mid^2=1$ leads to the same measurement statistics, i.e.
$\proj{i}\rho^\prime\proj{i}=\proj{i}\rho\proj{i}, \forall i$.
Obviously, the same holds for operators acting on $B$'s system.
Thus, if the measurement basis is known, we can choose any of
those reducible operators. If furthermore $A$ and $B$ symmetrize
their qubit pairs by the operations described in (II) and (III),
then the state describing their qubits has the form of Eq.
(\ref{rhogen}). If we then provide $E$ with a purification of this
state then we might only increase her power \footnote{We say that
$E$ has a purification of the state $\rho_{AB}$ if the state
describing $A$'s, $B$'s, and $E$'s system is $\ket{\Psi}_{ABE}$
such that $\rho_{AB}=\tr_E(P_{\ket{\Psi_{ABE}}})$.}. Note that the
symmetrization described in (II) and (III) commute with a
measurement in the $z$--basis.


In order to analyze the classical part of the protocol we
partially use some of the information--theoretic arguments
\cite{RenKoe04,KoMaRe03}, which have first been proposed in
\cite{ChReEk04} in order to analyze security of a large class of
QKD protocols \footnote{The proof technique introduced
in~\cite{ChReEk04} is based on the result of~\cite{KoMaRe03} and
the fact that the rank of a purification of $A$'s and $B$'s system
can be bounded.}. We assume that $A$ and $B$ hold strings $X^n$
and $Y^n$, resp., obtained by measuring a given state $\rho^n_{A
B}$, e.g., the state presented in~\eqref{rhogen}.

Let us first consider the CPP consisting of three steps. The
protocol is one-way, i.e., only communication from, say $A$ to $B$
is needed. I) \emph{Pre-processing:} Using her bit string $X^n$,
$A$ computes two strings $U^n$ and $V^n$, according to given
conditional probability distributions $P_{U|X}$ and $P_{V|U}$,
resp.  She keeps $U^n$ and sends $V^n$ to $B$. II)
\emph{Information reconciliation:} $A$ computes error correcting
information $W$ from $U^n$ and sends $W$ to $B$ \cite{footnote14}.
Using his information, $Y^n$ and $W$, $B$ computes a guess
$\hat{U}^n$ for $U^n$. III) \emph{Privacy amplification:} $A$
randomly chooses a function $F$ from a family of two-universal
hash functions and sends a description of $F$ to $B$
\cite{footnote12}. Then $A$ and $B$ compute their keys, $S_A =
F(U^n)$ and $S_B = F(\hat{U}^n)$, resp..

Let us introduce some notation before analyzing this protocol. We
describe the classical information of $A$ and $B$ as well as the
quantum information of $E$ by a tripartite density operator
$\rho_{X Y E}$ of the form $
  \rho^n_{X Y E}
=
  \sum_{x, y} P_{X^n Y^n}(x,y)
    P_{\ket{x}} \otimes P_{\ket{y}} \otimes \rho^{x,y}_E
$ where $\{\ket{x}\}_x$ and $\{\ket{y}\}_y$ are families of
orthonormal vectors and where $\rho^{x,y}_E$ is the quantum state
of $E$ given that $A$ and $B$'s values are $x$ and $y$, resp..
Similarly, $\rho_{S_A S_B E'}$ describes the classical key pair
$(S_A, S_B)$ together with the adversary's information $\rho_{E'}$
after the protocol execution.  We say that $(S_A, S_B)$ is
\emph{$\eps$-secure} if $ | \rho_{S_A S_B E'} - \sum_{s \in \cS}
P_{\ket{s}} \otimes P_{\ket{s}} \otimes \rho_{E'}| \leq \eps $.
Note that this definition leads to the so-called \emph{universally
composable} security, which implies that the key can safely be
used in \emph{any} arbitrary context \cite{RenKoe04}.

To determine the number $\ell^\eps_n$ of $\eps$-secure key bits
that can be generated by the above protocol, we use the following
recent results: I) The amount of key that can be extracted from a
string $U^n$ is given by the uncertainty of the adversary about
$U^n$, measured in terms of the so-called \emph{smooth R\'enyi
entropy}, $S_2^{\eps'}, S_0^{\eps'}$ ~\cite{RenKoe04}, as
introduced in~\cite{RenWol04}. II) The amount of information $B$
needs to correct his errors, using optimal error correction, is
given by his uncertainty about $A$'s string (again measured in
terms of the smooth R\'enyi entropy). Combining those results we
find for the number of $\eps$-- secure bits \cite{ReKr04},
\begin{equation}
  \ell_n^\eps
\approx
  \sup_{V^n \leftarrow U^n \leftarrow X^n} \bigl(
    S_2^{\eps'}(\rho^n_{U E V}) - S_0^{\eps'}(\rho^n_{E V})
    - H_0^\eps(U^n|Y^n V^n)
  \bigr)  \ ,\nonumber
\end{equation}
where ``$\approx$'' means that equality holds up to some small
term independent of $n$. In this formula, $\rho^n_{U E V }$ is the
density operator describing the string $U^n$ together with the
adversary's knowledge \cite{footnote13}. The supremum is taken
over all preprocessing applied by $A$.

In the remaining part of this paper we show how a lower bound on
the secret rate, $r:=\lim_{n \to \infty} (\ell^\eps_n/n)$, can be
determined considering only two--qubit density operators. To this
aim we first of all fix some pre--processing by $A$. We assume
that it is bit--wise, i.e. for each bit value $X_i$ she computes
$U_i$ and $V_i$ \footnote{A generalization to a pre--processing
where more bits are used is straightforward. Note however, that a
bitwise processing of the string $X^n$ might not be optimal.}. At
the end we take the supremum with respect to all those
pre--processing.

$A$ and $B$ symmetrize their qubits pairs by applying a random
permutation to the state $\rho_0$. Now we can assume, without loss
of generality, that the first $n_{p.e.}$  qubits are used for the
parameter estimation and the rest, $n_{data}$, is used to generate
the key. $A$ and $B$ estimate the error by measuring the
$n_{p.e.}$ qubits in all the different bases used by the protocol,
e.g. for the BB84, they measure in the $z$- and $x$-basis. Since
the state is symmetric and $n_{p.e}$ is sufficiently large, the
data qubits, which can then all be measured in the same basis, say
in the $z$-basis, contain the same amount of error. As explained
above, one can assume that the state describing the data qubits
has the simple form as in Eq. (\ref{rhogen}) \footnote{Note that
in principle $A$ and $B$ still have to symmetrize their qubit
pairs with respect to $\sigma_x\otimes \sigma_x$, however they can
do that after the measurement.}. Since the only free parameters
are the diagonal elements $\lambda_{n_1,n_2,n_3,n_4}$ (see Eq.
(\ref{rhogen})), the outcome of the parameter estimation implies
very strong conditions on them. In fact, conditioned on this
outcome the data qubits can be described by some state $\rho^n_{|
Q}$, where $Q=(n_1,n_2,n_3,n_4)/n$ is the frequency distribution
(depending on the parameter estimation outcome) of a
Bell--measurement. The state $\rho^n_{| Q}$ has the same structure
as the product state $\sigma_Q^{\otimes n}$, where $\sigma_Q$ is a
two--qubit Bell--diagonal state with eigenvalues $Q$. Due to this
similarity one can show that the smooth R\'enyi entropies of those
states are the same. Finally, using the fact that the smooth
R\'enyi entropy of a product state is asymptotically equal to the
von Neumann entropy~\cite{RenWol04}, we obtain the following lower
bound on the secret rate \cite{ReKr04}
\begin{equation} \label{eq:singlebound}
  r\geq
  \sup_{\substack{U \leftarrow X \\ V \leftarrow U}} 
    \, \inf_{\sigma_{A B} \in \Gamma_\QBER}
    \bigl(S(U | V E) - H(U|Y V) \bigr)
  \ .
\end{equation}
In this formula, $S(U | V E)$ denotes the von Neumann entropy of
$U$ conditioned on $V$ and $E$, i.e., $S(U|V E) := S(\sigma_{U V
E}) - S(\sigma_{V E})$. The state $\sigma_{U V E}$ is obtained
from $\sigma_{A B}$ by taking a purification $\sigma_{A B E}$ of
the Bell diagonal state $\cD_2(\sigma_{A
  B})$ and applying the measurement of $A$ followed by the
classical channels $U \leftarrow X$ and $V \leftarrow U$.
Similarly, $Y$ is the outcome of $B$'s measurement applied to the
second subsystem of $\sigma_{A B E}$. The set $\Gamma_\QBER$
contains all two--qubit states, $\sigma$, for which the protocol
computes a secret key when starting with the state
$\sigma^{\otimes n}$, where $\sigma$ is any state that $A$ and $B$
might share after a collective attack by $E$. Thus, in order to
prove full security for this class of QKD protocols one only has
to consider collective attacks. Note that, in order to compute a
lower bound $V$ can be discarded, however, the pre--processing
$X\rightarrow U$ turns out to be very important.

In order to derive this bound we assume that Eve has a
purification of the state $\sigma$. This is always possible as
long as the encoding/decoding operators ($A_{j}, B_{j}$) are
unitary. This implies that, for instance for the BB84 and the
6--state protocol coherent attacks are not more powerful than
collective attacks \footnote{Note that if we would consider a
pre--processing involving more than one bit, then also the states
in the generalized expression of ~\eqref{eq:singlebound} would be
higher dimensional (corresponding to this number of bits).}.

To reduce the number of parameters even further one might consider
only the set $\cD_2[\cD_1(\Gamma_\QBER)]$. It contains only
normalized two-qubit Bell--diagonal states, i.e. Eq.
(\ref{rhogen}) for $n=1$. Due to the fact that this state is
measured in the $z$-basis by $A$ and $B$ (and so is the $QBER=Q$)
we have $\lambda_1=1-Q-\lambda_2, \lambda_4=Q-\lambda_3$. The
considered protocol, i.e. the map ${\cal D}_1$, implies then
additional conditions on those coefficients.

Using techniques from quantum information theory, one can show
that if the supremum on the r.h.s. is also taken over any quantum
state $\rho_{U V}$ computed from $X$, then it is also an upper
bound for the rate $r$, i.e., $r \leq \min_\rho \sup_{V \leftarrow
U \leftarrow X} [ S(\rho_{U E V}) - S(\rho_{E V}) - H(U|V Y)]$,
where the minimum is taken over all states $\rho=\rho_{A B E}$
that can be generated by an attack of $E$ \cite{Wi}.

The case of individual attacks ($n=1$) has been widely studied,
using a bound (sometimes called Csisz\'ar and K\"orner bound)
which is similar to ~\eqref{eq:singlebound}, but without the extra
preprocessing terms: $X\rightarrow U \rightarrow V$. A priori, one
might think that the preprocessing $X \rightarrow U$ could not be
of any help, since the only choice $A$ has is to flip each bit
value with some probability, i.e. to introduce noise. However,
this noise differs clearly from the channel's noise. Although it
diminishes $A$'s mutual information with $B$ it may more severely
penalize $E$. For instance, for the 6-state protocol numerical
optimization shows that for all non zero QBERs it is always
advantageous for $A$ to first add some noise to her data, before
the EC and PA.

Let us now illustrate our result for several protocols. For the
BB84 the encoding/decoding operators are $A_1=B_1=V_x,
A_2=B_2=\one$, where $V_x$ is the Hadamard transformation. It is
easy to verify that ${\cal D}_2[{\cal
  D}_1(\rho_0)] =(1-Q-\lambda_1) P_{\ket{\Phi^+}}+\lambda_1
P_{\ket{\Phi^-}}+\lambda_1 P_{\ket{\Psi^+}}+(Q-\lambda_1)
P_{\ket{\Psi^-}}$ with $0\leq \lambda_1\leq Q$. After minimizing
the lower bound on the secret key rate (Eq.(\ref{eq:singlebound}))
with respect to $\lambda_1$, we optimize over the pre-processing
by $A$. We find for the optimal values $\lambda_1=Q-Q^2$ and
$q\rightarrow 0.5$, the probability for $A$ to flip the bit value,
that the secret key rate is positive for all $Q \leq 0.124$. Note
that if we would not optimize over the pre-processing by $A$, we
would obtain the well--known bound $0.1100$
\cite{ShPr00,ChReEk04}. Since the state $A$ and $B$ share, before
the EC and PA, is separable any entanglement based proof of
security fails. For the upper bound we obtain the known result
that the protocol is not secure if the QBER is higher than $0.146$
\cite{FuGr97}. For the $6$--state protocol we find that the secret
key rate is positive as long as $Q< 0.1412$ (known result $0.127$
\cite{Lo}). On the other hand, the protocol is insecure for all
$Q\geq 0.1623$. For the B92 we find a positive rate as long as
$\delta \leq 0.0278$ (known result $\delta \leq 0.0240$), where
$\delta$ characterizes the depolarization of a channel introducing
the same amount of noise ~\cite{ChReEk04,TaKo03}.


To conclude, we studied the security of a class of QKD protocols,
including BB84, $6$--state, B92 protocol among many others. We
presented a new security proof not based on entanglement
purification for all those protocols using one--way CPP. We show
that in order to prove full security one only has to consider
collective attacks. We derived a lower bound on the secret key
rate involving only entropies of two--qubit density operators. It
is shown that $A$ should add noise before the EC and PA phase.
Actually, this is why better bounds are achieved and also the
reason why entanglement based proofs would fail here. We
illustrated our results by presenting new bounds on all the
protocols mentioned above.

N. G. and B. K. thank the Swiss NCCR ''Quantum photonics'' and the
European IST project SECOQC.


\begin{thebibliography}{99}
\bibitem{footnote2} We consider here only qubits, however a generalization to
higher--dimensional systems is straightforward.
\bibitem{footnote6} The interaction has to be unitary since $E$ is restricted to the laws of quantum
physics.
\bibitem{footnote61} Note that for the BB84, or the $6$--state protocol, these filtering operations would be just
unitaries.
\bibitem{footnote0} ${\cal U}_{BE}$ might no longer be
unitary here since we consider the state after the sifting,
however, this will not change the following arguments.
\bibitem{footnote7}
We distinguish between I)Individual attacks: ${\cal
U}_{EB}=U_1^{\otimes n}$, and indepenent measurement of auxiliary
systems right after $E$ knows ${\bold j}$, II) Collective attacks:
${\cal U}_{EB}=U_1^{\otimes n}$, and collective measurement at the
very end of the protocol, including the CPP, III) Coherent
attacks: general unitary and general measurement at the very end
of the protocol.
\bibitem{footnote31} $E$ might be able to replace the channel
with an ideal channel. However, it is reasonable to assume that
the detector noise is not under $E$'s control. Thus, we only
consider here the situation of a noisy channel.
\bibitem{footnote10} By equivalent we mean that we do not change the protocol, in the
sense that we keep the sets ${\cal S}_0$, or ${\cal S}_1$ and the
information of $A$, $B$ and $E$ is not changed.
\bibitem{footnote9} We use the notation
$P_{\ket{\Phi}}=\proj{\Phi}$, for any state $\ket{\Phi}$.
\bibitem{footnote14} For instance, $W$ can be the output of a two-universal
  function applied to $U^n$.
\bibitem{footnote13} Since $V^n$ is sent from $A$ to $B$ over an insecure channel, $E$ might
  know these values as well.
\end{thebibliography}
\end{document}